\documentclass{article}
\usepackage{amsmath}
\usepackage[preprint]{neurips_2026}

\usepackage{longtable}
\usepackage[utf8]{inputenc} 
\usepackage[T1]{fontenc}    
\usepackage{hyperref}       
\usepackage{url}            
\usepackage{booktabs}       
\usepackage{graphicx}       
\usepackage{amsfonts}       
\usepackage{nicefrac}       
\usepackage{microtype}      
\usepackage{xcolor}         

\usepackage{tabularx}
\usepackage{makecell}

\title{
Beyond Neural Activity Prediction: Probing Latent Representations in Mouse V1 Digital Twins
}

\author{%
    Adriano Lima\thanks{Equal contribution.} \\
    Department of Psychological \& Brain Sciences \\
    University of California, Santa Barbara \\
    Santa Barbara, CA 93117 \\
    \texttt{adrianolimacotralhaa@ucsb.edu}
  \And
    Yuchen Hou\footnotemark[1] \\
    Department of Computer Science \\
    University of California, Santa Barbara \\
    Santa Barbara, CA 93117 \\
    \texttt{yuchenhou@ucsb.edu}
  \And
    Michael Beyeler \\
    Department of Computer Science \\
    Department of Psychological \& Brain Sciences \\
    University of California, Santa Barbara \\
    Santa Barbara, CA 93117 \\
  \And  
    Marius Schneider \\
    Department of Bioengineering\\
    University of California, Santa Barbara \\
    Santa Barbara, CA 93117 \\
    \texttt{marius\_schneider@ucsb.edu}
}

\begin{document}

\maketitle

\begin{abstract}
Digital twins of sensory cortex serve as powerful response oracles for predicting neural activity to novel stimuli. Although prediction accuracy is the central metric by which these models are evaluated, it provides limited insight into the latent representations that support those predictions. This limitation becomes increasingly important as digital twins are used as in silico experimental systems for stimulus design and hypothesis generation: models with similar prediction accuracy may rely on different latent representations. We address this gap by systematically probing a family of convolutional-recurrent digital twins of mouse V1 trained to predict neural activity from naturalistic videos recorded in freely moving mice. The models share the same training data and neural-prediction objective, but differ in visual-encoder architecture. For each frozen model, we characterize latent representations along three levels: (i) linear decodability from controlled visual probes of orientation, contrast, and motion; (ii) latent-unit tuning to canonical visual features including orientation selectivity, contrast response, spatial-frequency tuning, and phase sensitivity; and (iii) population geometry of hidden-layer activity. 
Across architectures, better neural-response prediction correlates with stronger probe accuracy. Additionally, highly predictive models exhibit flatter hidden-population eigenspectra, indicating more distributed, higher-dimensional representations closer to population-geometry signatures reported in mouse V1.
Although these representational properties covary with prediction accuracy across architectures, digital twins with comparable prediction scores can still differ substantially in probe performance and latent-unit tuning. These results establish multi-level representational probing as a complement to standard neural-prediction evaluation, providing a framework for understanding digital twins not only as predictors, but also as substrates for studying visual computations.
\end{abstract}

\section{Introduction}
Digital twins of sensory cortex are emerging as powerful response oracles. Trained on large-scale neural recordings, these models predict neural responses to novel visual stimuli and can support in silico stimulus design, model comparison, and hypothesis generation. In mouse primary visual cortex (V1), data-driven response models have progressed from convolutional system-identification models to larger multimodal architectures trained on natural images and naturalistic videos \citep{klindt2017whatwhere,lurz2021generalization,xu2023multimodal,turishcheva2024dynamic,wang2025foundation}. Community benchmarks such as Sensorium have accelerated this progress by comparing models primarily through held-out neural-prediction metrics, including correlations between predicted and recorded responses \citep{willeke2022sensorium,turishcheva2024reproducibility,schneider_mouse_2025}. 
Such prediction-based evaluation is necessary: a digital twin that does not predict neural activity cannot serve as a response oracle.
However, output fidelity alone does not specify what visual information the model represents, how that information is organized, or whether different high-performing models arrive at similar latent structure.

This limitation reflects a broader underdetermination problem in neural-response modeling. Multiple architectures can fit similar stimulus-response mappings while relying on different latent representations (e.g. \citet{conwell2024largescale}). Such differences become central when digital twins are used as in silico experimental systems to generate hypotheses about biological visual cortex or draw conclusions about sensory computation. In that setting, prediction accuracy is a necessary but incomplete evaluation criterion: we also need to know what visual computations are accessible in the model's latent states, how individual latent units are tuned, and how population activity is organized. 
Importantly, while digital twins are not circuit-level mechanistic models of cortex, they are fully accessible: every hidden unit, layer, and population response can be measured under arbitrary stimuli.

A broad literature comparing networks trained on common computer vision objectives to visual cortex has shown that objective, architecture, and training data can affect how well model representations align with neural measurements \citep{schrimpf2018brainscore,cadena2019howwell,shi2019comparison,nayebi2023mouse,conwell2024largescale,cadena2024diverse}. 
In parallel, recent studies have begun to use data-driven neural prediction models as in silico experimental systems rather than only as response predictors. Digital twin models have been used to investigate phenomena beyond held-out neural-response prediction, including stimulus invariances, functional cell-type structure, population geometry, and hierarchical object representations \citep{ding2026bipartite,burg2024discriminative,liscai2025beyond,luconi2025anatomically}. Together, these studies suggest that targeted analyses of trained digital twins can reveal functional and representational structure not fully captured by prediction metrics alone.

However, there is no systematic evaluation approach to investigate what visual computations are linearly accessible, what canonical tuning properties emerge in latent units, and how hidden population activity is geometrically organized in digital twins of mouse V1.
Here, we introduce a multi-level evaluation framework for probing latent representations in digital twins. We apply this framework to a family of convolutional-recurrent models trained to predict V1 activity from naturalistic video recorded in freely moving mice. The models share the same training data and neural-prediction objective but differ in their visual-encoder architecture components.
This architecture sweep provides a controlled setting for systematically analyzing how visual representations are learned, testing whether improvements in neural prediction are accompanied by systematic changes in representational organization, and whether representational differences persist despite comparable output fidelity. For each frozen model, we evaluate three complementary levels: 
(i) linear-probe accuracy on controlled orientation, contrast, and motion-direction probes;
(ii) latent-unit tuning to canonical visual dimensions (orientation selectivity, contrast response, spatial-frequency tuning, phase sensitivity); 
(iii) population geometry of hidden activity, summarized by the GRU eigenspectrum.
These levels were chosen to connect model evaluation with interpretable descriptors from systems neuroscience and representation learning.

\begin{figure*}[ht!]
\centering
\includegraphics[width=\textwidth,keepaspectratio=true]{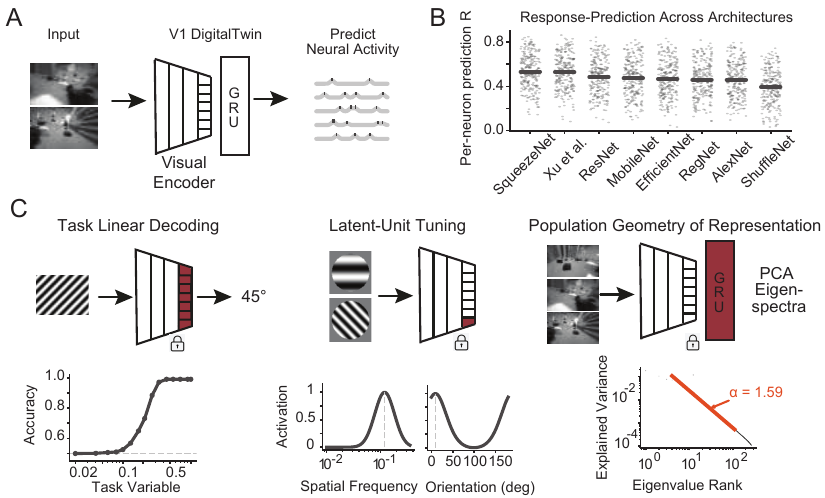}
    \caption[Caption]{{Multi-level probing of hidden representations in V1 digital twins.}
    \textbf{(A)} We compare V1 digital twins trained to predict neural activity from videos recorded in freely moving mice. Models share the same neural prediction objective and differ in the visual-encoder architecture.
    \textbf{(B)} Held-out neural prediction performance across visual-encoder architectures. Gray points indicate per-neuron prediction correlations and horizontal bars indicate architecture means.
    %
    \textbf{(C)} We characterize latent representations along three levels. Left, we measure \emph{task decodability} by training linear readouts on controlled visual probes targeting orientation, contrast, and motion information. Middle, we estimate \emph{latent-unit tuning} by presenting parametric stimuli and measuring responses of individual hidden units to canonical visual dimensions. Right, we quantify \emph{population geometry} by computing the eigenspectrum of hidden-layer activity and fitting a power-law exponent $\alpha$, where $\lambda_k \propto k^{-\alpha}$.
    }
     \label{fig1}
\end{figure*}

Across architectures, we find that neural-prediction accuracy is informative but does not fully characterize a digital twin's internal representations.
Better response predictors tend to support stronger linear readouts of orientation, contrast, and motion detection tasks, and exhibit flatter hidden-state eigenspectra closer to population-geometry signatures reported for mouse V1. Latent-unit analyses further show that response-predictive training gives rise to interpretable tuning along classical visual dimensions. 
Crucially, however, prediction accuracy does not uniquely determine representational organization: models with similar neural performance can differ in task accuracy, latent-unit tuning, and population geometry.

Our contributions are: First, we systematically probe the latent representations of convolutional–recurrent digital twins of mouse V1 along three levels: linear-probe task readouts, latent-unit tuning, and population geometry, using twins trained for neural prediction. 
%
%
Second, we show that several representational properties (linear decodability of low-level visual information, classical tuning structure, and population-geometry signatures of mouse V1) emerge with the prediction objective and covary with prediction quality across architectures. 
Third, we establish multi-level representational probing as a complement to neural prediction metrics, providing an evaluation strategy for comparing digital twins not only by how accurately they predict neural responses, but also by what their latent representations make available for in silico analysis.

\section{Methods}
\label{sec:methods}
We evaluated whether neural-prediction models of mouse primary visual cortex learn representations that support mouse-relevant visual computations. The analysis pipeline contained five substantive components, with frozen feature extraction serving as the bridge from trained neural-prediction models to downstream probes.
First, we trained a family of multi-mouse V1 neural-prediction models based on the freely moving digital-twin architecture of \citet{xu2023multimodal}. 
The recurrent component was held fixed while the visual encoder was varied
(\S\ref{sec:methods:model_training}). Second, we froze each trained model and extracted fixed representations for downstream analysis
(\S\ref{sec:methods:feature_extraction}). Third, we trained linear probes on synthetic visual tasks designed to measure task-relevant information in the learned representation (\S\ref{sec:methods:behavioral_readouts}). Fourth, we characterized emergent feature tuning using parametric gratings and quantified hidden representation geometry using PCA eigenspectra
(\S\ref{sec:methods:emergent_tuning}). Finally, we related neural-prediction performance, behavioral readout performance, population geometry, and tuning summaries across models using cross-model scatter and regression analyses (Appendix~\ref{app:cross_model_stats_details}). Detailed architecture, training, task, feature-extraction, tuning, and statistical-analysis settings are provided in Appendix~\ref{app:methods}.

\subsection{Neural-prediction training and frozen feature extraction}
\label{sec:methods:model_training}
\label{sec:methods:feature_extraction}

\paragraph{Model family and encoder architectures.}
We trained a family of convolutional-recurrent neural-prediction models on
mouse V1 activity. Each model mapped grayscale movie frames through a visual encoder, integrated the resulting feature sequence with a single-layer GRU, and predicted neural activity with mouse-specific readout heads. The visual encoder and GRU were shared across mice, whereas the final readout heads were mouse-specific. The primary model variation was the visual encoder family, while recurrent integration, mouse-specific readout structure, and the training objective were held fixed to systematically analyze how visual representations learned under neural-prediction training. The encoder family comprised CNN, ResNet-18, EfficientNet-B0, MobileNet-V2, AlexNet, SqueezeNet-1.1, ShuffleNet-V2, and
RegNet-Y-400MF; detailed architecture settings are provided in
Appendix~\ref{app:model_zoo_architectures}.

\paragraph{Neural data, training, and frozen features.}
Models were trained jointly on publicly available neural-response recordings from three mice freely running in an arena (see \citet{parker_joint_2022}). Each session was represented as temporally ordered movie frames paired with simultaneous V1 single-neuron activity. Training examples were five-frame grayscale clips with a 48 ms sampling rate, and each target was the neural response at the next time bin. Each encoder architecture was trained across three run seeds, and for each architecture-seed pair we retained the checkpoint with the best validation objective. 

A task-specific linear probe was trained on frozen model latent features for downstream representational analyses. Feature
extraction was matched to the temporal structure of each task.
Static orientation discrimination task used $108$-dimensional visual-encoder outputs from single-frame stimuli. Dynamic Contrast and Motion RDK tasks used $512$-dimensional GRU hidden-state features after recurrent temporal integration. Linear probes were trained on these frozen features using matched synthetic stimulus draws across architectures within each run seed. 

\subsection{Behavioral readout tasks}
\label{sec:methods:behavioral_readouts}
The behavioral readout suite probed three synthetic visual computations not used in neural-prediction training: static grating orientation, flashed-target
contrast detection, and coherent-motion direction discrimination. See \ref{app:behavioral_readout_details} for details.

Orientation task was an eight-way static grating classification task. Each trial was a single full-field sinusoidal grating with orientation sampled from one of eight bins spanning $[0^\circ,180^\circ)$. Performance was classification accuracy over a training-size sweep, with accuracy at the largest training-set size used as the scalar endpoint.

Dynamic Contrast task was a binary target-detection task using five-frame clips. On positive trials, a Gabor target was flashed in a fixed visual-field patch; on negative trials, the clip contained only noisy background. Probes were evaluated across target contrasts from blank or near-blank stimuli through high contrast. Performance at each contrast was summarized by balanced accuracy, and the primary endpoint was the area under the balanced-accuracy curve.

Motion RDK task was a two-alternative direction-discrimination task using $80$-frame random-dot kinematogram clips. Probes were trained on high-coherence conditions and evaluated on a coherence grid from chance-level motion through high-coherence stimuli. Clips were processed in five-frame chunks, and the final chunk's GRU hidden state was used as the clip-level feature. The primary endpoint was the area under the accuracy-versus-coherence curve.

\subsection{Emergent tuning and representation geometry}
\label{sec:methods:emergent_tuning}

We measured latent tuning at the final visual-encoder output by presenting
static full-field sinusoidal gratings to each frozen encoder. Treating each output dimension as a latent feature unit, we summarized orientation selectivity, phase modulation, spatial-frequency preference, and signed contrast-response tuning for cross-model analyses. In addition, we quantified latent representation geometry using the eigenspectrum of GRU hidden activity. For each trained model, we extracted final GRU hidden states from 5{,}000 clips of a fixed reference mouse session, computed PCA eigenspectra, and fit a power-law exponent over ranks 20-200. The resulting GRU eigenspectrum exponent was used as the representation-geometry summary in the cross-model analyses.


\section{Results}

\subsection{A multi-level framework for probing V1 digital-twin representations}

Digital twins are typically optimized and evaluated as neural response predictors.
This output-level evaluation does not specify what information is represented in the model's latent states. To characterize these learned representations directly, we evaluated digital twins along multiple levels alongside the neural prediction accuracy.

We trained eight convolutional-recurrent models to predict V1 activity from naturalistic videos recorded in freely moving mice (Fig. \ref{fig1}A). 
Across architectures, the models achieved comparable held-out prediction performance, establishing that each model could serve as a response oracle while also exhibiting measurable differences in output fidelity (Fig. \ref{fig1}B).

We then probed the latent representation of the tested visual encoders along three levels (Fig. \ref{fig1}C). First, we measured  \emph{linear functional access}: whether information about controlled visual variables could be recovered from frozen latent features using simple linear readouts. 
%
%
We used three probes targeting orientation, contrast, and motion information to provide controlled assays of visual computations that are classically relevant for mouse V1 and can be compared across digital twins.

Second, we quantified \emph{latent-unit tuning}. Using parametric stimuli, we measured responses of individual latent units to orientation selectivity, contrast response, spatial frequency tuning, and phase sensitivity. This analysis investigates whether the models’ individual units develop interpretable selectivity along visual dimensions commonly used to characterize V1 responses.

Third, we measured \emph{population-level geometry} of hidden-layer activity. For each model, we computed the eigenspectrum of population responses to naturalistic stimuli and fit a power-law exponent $\alpha$ to summarize how variance is distributed across population dimensions. This provides a global descriptor of representational organization that complements the unit-level tuning analysis. 
These three levels allow us to separate output fidelity from representational properties, which will be analyzed in detail in the following sections.

\subsection{Controlled visual probes reveal linear functional access in V1 digital twins}
\begin{figure}[t]
    \centering
    \includegraphics[width=1\linewidth]{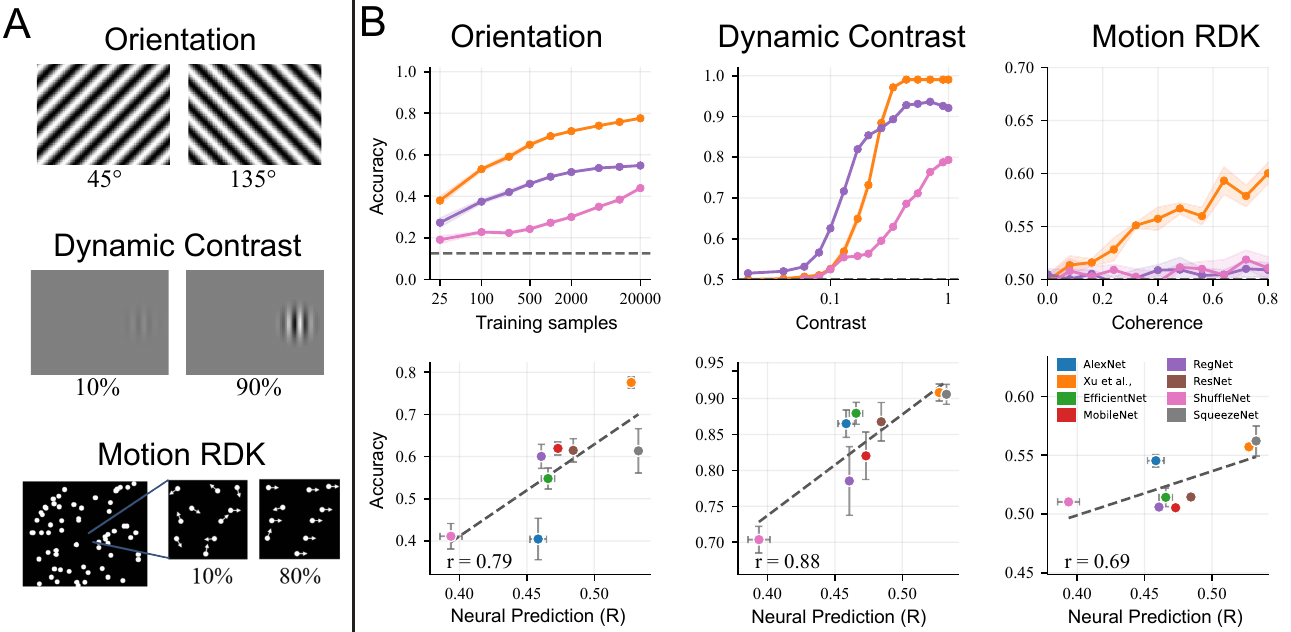}
    \caption{\textbf{Controlled visual probes reveal functional access in V1 digital twins.} 
    \textbf{(A)} Example stimuli for the three controlled visual probes: orientation discrimination, dynamic contrast detection, and RDK motion-direction discrimination.
    \textbf{(B)} Linear readout performance across visual-encoder architectures. Top: representative task-performance curves for each probe. Bottom: architecture-level relationships between probe performance and held-out neural prediction accuracy. Points indicate architecture means with seed variability; correlations are computed across architecture means.
    } \label{fig:stimuli}
\end{figure}

\paragraph{Linear readouts recover task-relevant information across architectures.}
To assess what visual information is linearly decodable from the latent representations of V1 digital twins, we froze each model's latent features and trained simple linear readouts on three controlled visual probes targeting orientation, dynamic contrast, and motion information (Fig.~\ref{fig:stimuli}A, S. Table~\ref{tab:model_performance}). 

All architectures supported non-trivial linear readout performance on two of the three probes, but the level of accessible information differed substantially across models (Fig.~\ref{fig:stimuli}B, top). For the orientation probe, readout accuracy was considerably above chance even at very small training sample sizes, indicating that frozen latent representations contain recoverable information about orientation. However, the performance of this readout varied across architectures, with some models reaching substantially higher accuracy than others (e.g., $0.411 \pm 0.052$ in ShuffleNet vs. $0.776 \pm 0.022$ in \cite{xu2023multimodal}). 
For the dynamic contrast probe, all models exhibited psychometric-like performance curves, with balanced accuracy increasing as contrast increased, but again with clear differences in slope and saturation across architectures. 
For the motion-direction probe based on random-dot kinematograms (RDKs), performance remained closer to chance overall, indicating that motion information was less accessible than orientation or contrast in these frozen representations. Nonetheless, a subset of models  (e.g., $0.562 \pm 0.022$ in SqueezeNet and $0.545 \pm 0.009$ in AlexNet) showed clearly elevated, coherence-dependent readout while most remained close to chance, indicating that direction-discrimination accessibility is not uniform across the model family.
\paragraph{Task performance covaries with neural prediction across architectures.}
To summarize these differences, we computed a scalar performance metric for each probe and related it to held-out neural prediction accuracy across architectures (Fig.~\ref{fig:stimuli}B, bottom). 
For orientation, we used out-of-distribution accuracy at the largest training-set size; for dynamic contrast, balanced-accuracy AUC across contrast levels; and for RDKs, accuracy AUC across motion coherence. Across architecture means ($n=8$), models with better neural prediction tended to also support stronger linear readouts on all three probes. This relationship was strongest for dynamic contrast detection ($r=0.88$, $p=0.0037$) and orientation discrimination ($r=0.79$, $p=0.0209$), and weaker for motion-direction discrimination ($r=0.69$, $p=0.0606$), where the trend was preserved but marginally below significance. 

However, architectures with comparable neural-prediction performance can occupy different positions on each task readout. For example, both \citet{xu2023multimodal} and SqueezeNet have similar neural prediction scores ($0.527 \pm 0.154$ and $0.533 \pm 0.157$), and perform similarly in dynamic contrast ($0.908 \pm 0.020$ and $0.906 \pm 0.024$), but have substantially different orientation-probe accuracy ($0.776 \pm 0.022$ and $0.614 \pm 0.091$). This discrepancy suggests that prediction accuracy reflects, but does not fully determine, the functional structure of twin representations.
\begin{figure}[t]
    \centering
    \includegraphics[width=1\linewidth]{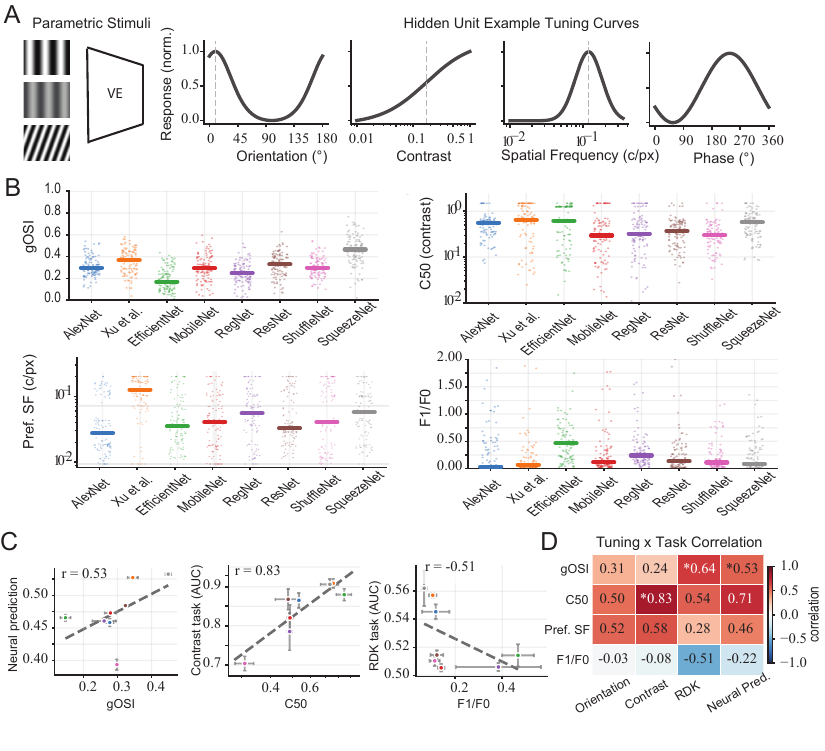}
    \caption{\textbf{Latent-unit tuning provides interpretable axes  for comparing V1 digital twins.} 
    \textbf{(A)} We probed latent units with parametric visual stimuli varying canonical stimulus dimensions, including orientation, contrast, spatial frequency, and phase. Example latent units show structured responses to orientation, contrast, and spatial frequency, from which we derived tuning metrics such as global orientation selectivity (gOSI), contrast semisaturation ($C_{50}$), preferred spatial frequency, and phase modulation (F1/F0).
    \textbf{(B)} Distributions of latent-unit tuning metrics across visual-encoder architectures. Points indicate individual hidden units and horizontal bars indicate architecture-level medians.
    \textbf{(C)} Selected architecture-level relationships between tuning metrics and neural prediction or functional readout performance. Points indicate architecture means with seed variability shown where available; correlations are computed across architecture means.
    \textbf{(D)} Architecture-level correlations between tuning metrics and controlled visual probes or neural prediction performance. 
     * p<0.05. Cluster Robust OLS on seed-level data, clustering by architecture (n=8 architectures, 3 random seeds each). 
} 
    \label{fig3}
\end{figure}

\subsection{Latent-unit tuning provides interpretable but partial axes of functional access}
%
The controlled probes in Fig.~\ref{fig:stimuli} show that the visual information accessible from frozen digital-twin representations varies across architectures. We then tested whether these differences are reflected in the tuning properties of individual latent units. Using parametric visual stimuli, we measured tuning to canonical visual dimensions commonly used to characterize V1 responses \citep{niell2008}. More specifically, for each architecture, we summarized latent-unit responses using global orientation selectivity (gOSI), contrast-response parameters (\(C_{50}\)), preferred spatial frequency, and phase sensitivity (F1/F0). \(C_{50}\) denotes the contrast at which the fitted response reaches half its saturating amplitude, and F1/F0 quantifies phase sensitivity, with larger values indicating stronger phase modulation and smaller values indicating more phase-invariant responses.

Latent units exhibited structured tuning along all measured visual dimensions (Fig.~\ref{fig3}A, B). Individual units showed orientation-selective responses, contrast-dependent response profiles, spatial-frequency preferences, and phase-dependent modulation. 
Across architectures, the distributions of these descriptors differed, indicating that models trained with the same neural-prediction objective can nevertheless develop distinct unit-level response profiles. 
However, most models shared moderate orientation selectivity ($0.297 \pm 0.085$) and low spatial-frequency preferences ($0.126 \pm 0.050$ cycle-per-pixel, equivalent to $0.084 \pm 0.033$ cycle-per-degree), which were broadly compatible with values reported for mouse V1 under grating stimulation~\citep{niell2008}. Note that we use these comparisons as biological reference ranges, not as evidence of one-to-one equivalence between hidden units and cortical neurons.

We then explored whether these tuning properties relate to task performance (Fig.~3C, D). Orientation-probe performance marginally correlated with global orientation selectivity gOSI ($r = 0.31, p = 0.053$). Contrast-probe performance was most strongly associated with the contrast-response midpoint $C_{50}$ ($r = 0.83, p=0.002$), indicating that C50 was the tuning property most directly relevant to contrast sensitivity. Motion-direction performance in the RDK probe was associated with gOSI ($r = 0.64, p = 0.006$) and showed a marginal negative correlation with the phase sensitivity index F1/F0 ($r = -0.51, p = 0.056$). These relationships suggest that latent-unit tuning provides interpretable information for understanding downstream visual tasks, and that no individual tuning statistic fully determines all task performance.

\begin{figure}[t]
    \centering
    \includegraphics[width=1\linewidth]{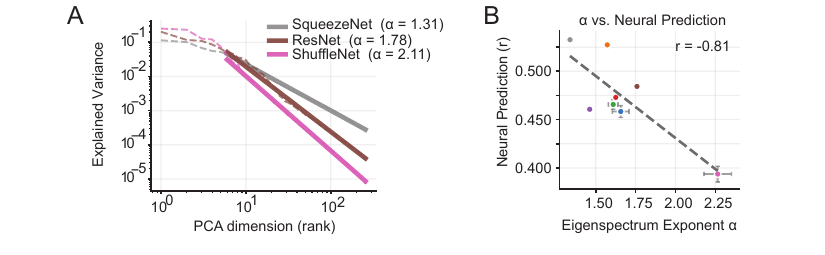}
    \caption{\textbf{Population geometry of hidden representations covaries with neural prediction performance.} 
    \textbf{(A)} Example eigenspectra of latent-layer activity for three visual-encoder architectures. For each model, hidden activations were collected across naturalistic video stimuli, decomposed with PCA, and summarized by the eigenspectrum of explained variance. Solid lines show power-law fits, $\lambda_k \propto k^{-\alpha}$, over the fitted rank range; lower $\alpha$ corresponds to a flatter spectrum with variance distributed across more population dimensions.
    \textbf{(B)} Eigenspectrum exponent $\alpha$ is negatively associated with held-out neural prediction performance across architectures. Points show architecture means and error bars indicate variability across seeds; the correlation is computed across architecture means.
    }
    \label{fig4}
\end{figure}

\subsection{Population geometry of hidden activity covaries with neural prediction accuracy}
To characterize organization beyond individual hidden units, we used the eigenspectrum of hidden-layer activity to quantify population geometry. For each architecture, we collected hidden states in response to naturalistic video inputs, computed the covariance matrix of the resulting population activity, and fit a power law, \(\lambda_k \propto k^{-\alpha}\), over a fixed range of principal-component ranks.  
The exponent $\alpha$ summarizes how variance decays across dimensions: larger $\alpha$ indicates activity dominated by a few dimensions, smaller $\alpha$ means a flatter, higher-dimensional representation. \citet{stringer2019high} showed that mouse V1 responses to high-dimensional natural images exhibit approximately power-law eigenspectra with exponents close to 1: high-dimensional enough to be expressive, yet structured enough to support smooth stimulus representations.

Recent work has applied this descriptor to digital twins at the level of predicted neural responses, revealing that twins can achieve high single-neuron prediction while failing to reproduce biological population geometry \citep{liscai2025beyond}. We instead apply the descriptor to the hidden representations that produce these predictions, to explore whether the latent population geometry varies systematically with prediction quality across architectures.

Overall, the fitted exponents were steeper than the values reported for mouse V1  (Fig.\ref{fig4}A), indicating that hidden activity remained more concentrated in leading dimensions than biological V1 population responses. However, architectures with better neural prediction accuracy tended to have lower \(\alpha\), corresponding to flatter and more distributed hidden population activity closer to population-geometry signatures reported for mouse V1 ($r = -0.81, p<0.001$, Fig.\ref{fig4}B). Among the representational descriptors we measured, \(\alpha\) showed one of the strongest associations with neural prediction performance.
Together, these results show that eigenspectrum analysis provides a population-level descriptor of hidden representations that complements both neural prediction accuracy and hidden-unit tuning.


\section{Discussion and Limitations}
Digital twins of sensory cortex are built to predict neural responses, and prediction accuracy is therefore the central metric for evaluating their success as response oracles. Here, we ask a complementary question: what hidden representations do these response oracles develop?
We characterized latent representations in a family of V1 digital twins trained with the same neural-prediction objective but different visual-encoder architectures. 
We evaluated their latent representations using three complementary probes: linear readouts measured which low-level visual variables were linearly accessible from frozen latent states, tuning analyses described how individual latent units responded to canonical stimulus dimensions, and eigenspectra summarized the population-level organization of hidden activity.
%
Because no task- or tuning-related objective was used during training, these analyses could reveal what representational structure emerges from neural-response prediction alone. 
Across the architectures tested, stronger neural prediction was associated with stronger readouts in orientation, contrast, and motion information, as well as flatter, more distributed population eigenspectra. Latent-unit tuning provided interpretable descriptors along classical visual dimensions and revealed architecture-dependent differences in unit-level response properties. Together, these results establish representational characterization as a complementary way to compare V1 digital twins beyond prediction accuracy alone.

\paragraph{Prediction accuracy underdetermines representation.}
The flexibility that makes modern digital twins effective response predictors also makes their internal organization underdetermined: high-capacity architectures can fit similar stimulus-response mappings while converging to different latent representations. This matters when digital twins are used as scientific instruments rather than only as response oracles, because two models with comparable prediction accuracy can fail on different stimulus regimes, support different downstream computations, or learn distinct intermediate representations of the same input. 
Unlike the biological circuits whose stimulus-response mapping they approximate, digital twins provide direct access to their internal states; every latent feature can be measured and perturbed under controlled stimulus conditions.
Representational probing therefore turns underdetermination from a limitation into a question that can be answered empirically: which representational differences persist among well-predicting twins, and which level do those differences covary along? 
Our results suggest that better prediction is accompanied by stronger functional readouts and more V1-like population geometry, while meaningful representational differences remain among models with similar prediction scores.

\paragraph{Population geometry in hidden representations.}
Among the three representational levels, population geometry showed the strongest association with prediction accuracy, with better predictors exhibiting lower $\alpha$ and more distributed hidden activity. This complements work on population geometry at the level of predicted neural responses: \citet{liscai2025beyond} found that output-stage geometry can diverge from biological V1 even when single-neuron prediction is strong, and that shifting output geometry toward biological V1 requires additional regularization and decreases prediction accuracy. At the level of hidden states, we observe that higher prediction accuracy was associated with hidden-population geometry shifted toward (though not reaching) the V1 regime, despite no explicit population-geometry objective.

\paragraph{Limitations.}
Our analysis covers eight visual-encoder architectures trained within the same convolutional-recurrent digital-twin framework, on the same freely moving mouse V1 dataset, and with the same neural-prediction objective. This design isolates architectural variation within one model family, but it does not cover the full design space of digital twins, including transformer-based or otherwise substantially different architectures. The cross-architecture relationships should therefore be interpreted as descriptive trends within the tested model family. A larger architectural sweep will be needed to determine whether the relationships between prediction accuracy, functional access, tuning, and population geometry generalize across broader classes of neural-response models.

\paragraph{Future directions.}
Our evaluation method can be used to address questions that are not captured by aggregate prediction scores. First, applying the probes across layers would reveal how visual information, tuning, and population geometry are built across the hierarchy of a digital twin. Second, applying the framework across model families, datasets, and training objectives would test whether different digital twins converge to distinct representations with similar output performance. Third, two twins with similar averaged scores may fail on different stimulus regimes or predict different subsets of neurons well. Testing whether the error profiles can be predicted from a model's representations would make hidden-state analysis directly useful for digital twin selection.

\paragraph{Conclusion.}
Digital twins are built to predict neural responses, and prediction accuracy remains the central measure of their value as response oracles. Our results show that models with the same neural-prediction objective can differ in controlled-probe performance, latent-unit tuning, and population geometry. These differences are related to prediction quality, but they are not summarized by a single prediction correlation score. The latent representations of digital twins therefore deserve direct evaluation: they reveal what visual structure emerges from neural-response prediction, and provide a basis for comparing digital twins beyond accuracy.

\paragraph{Data and Code Availability.} The neural recording data used to train and evaluate the digital twins is publicly available through \href{https://osf.io/msp3a/overview}{osf.io}. All code, including model training, analyses, and evaluation pipeline, will be publicly available upon publication. 

\newpage

\bibliographystyle{plainnat}
\bibliography{references}

\newpage

\clearpage
\appendix
\section{Detailed Methods}
\label{app:methods}

\subsection{Neural-prediction training details}
\label{app:model_training_details}

\paragraph{Mouse sessions.}
We used publicly available freely moving neural recordings from \citet{parker_joint_2022}, which are accessible through \href{https://osf.io/msp3a/overview}{osf.io/msp3a/overview}.   
The neural-prediction models were trained jointly on three mouse V1 sessions:
\texttt{070921\_J553RT}, \texttt{110421\_J569LT}, and
\texttt{101521\_J559NC}. Following \citet{xu2023multimodal}, for each session, data were organized into ten temporal
segments. Training used the session-specific training split from each segment,
with the first $80\%$ of each segment used for optimization and the remaining
$20\%$ used for validation. Test-split segments were held out for downstream
neural-prediction evaluation. In the current multi-mouse training pipeline, all
recorded units are retained for each mouse-specific readout.

\paragraph{Dataset construction.}
For a mouse $m$, each training example consisted of a movie snippet
$\mathbf{x}_{t:t+4}^{(m)}\in\mathbb{R}^{5\times1\times60\times80}$ and a
neural-response target $\mathbf{y}_{t+5}^{(m)}\in\mathbb{R}^{N_m}$, where $N_m$
is the number of recorded units for mouse $m$. Movie arrays were converted to
single-channel tensors and normalized to $[0,1]$ by subtracting the movie minimum
and dividing by the movie maximum when positive. The training dataloaders cycled
through the three mice and updated the shared visual encoder and GRU together
with the active mouse's readout head.

\paragraph{Architecture.}
Each frame was processed independently by the
shared visual encoder $f_\theta$, producing a sequence of feature vectors
\[
    \mathbf{z}_{t+i} = f_\theta(\mathbf{x}_{t+i}) \in \mathbb{R}^{108},
    \qquad i=0,\ldots,4.
\]
The sequence $(\mathbf{z}_t,\ldots,\mathbf{z}_{t+4})$ was then passed to a
single-layer GRU with hidden size $512$. The final hidden state
$\mathbf{h}_{t+4}\in\mathbb{R}^{512}$ was mapped to neural responses by a
mouse-specific linear readout $W_m\mathbf{h}_{t+4}+\mathbf{b}_m$ followed by a
softplus nonlinearity:
\[
    \widehat{\mathbf{y}}^{(m)}_{t+5}
    = \operatorname{softplus}(W_m\mathbf{h}_{t+4}+\mathbf{b}_m).
\]
The visual encoder and GRU parameters were shared across all mice, whereas the
readout matrices $W_m$ and biases $\mathbf{b}_m$ were separate for each mouse.

\paragraph{Loss.}
For each minibatch from mouse $m$, the primary loss was Poisson negative
log-likelihood with non-log predictions,
\[
    \mathcal{L}_{\mathrm{Poisson}}
    = \frac{1}{BN_m}\sum_{b=1}^{B}\sum_{j=1}^{N_m}
    \left(\widehat{y}_{bj}^{(m)} - y_{bj}^{(m)}\log \widehat{y}_{bj}^{(m)}\right),
\]
with the constant term omitted. We added an L1 penalty on the active mouse's
readout weights,
\[
    \mathcal{L}
    = \mathcal{L}_{\mathrm{Poisson}}
    + \lambda_{\mathrm{L1}}\frac{\lVert W_m\rVert_1}{|W_m|},
\]
where $\lambda_{\mathrm{L1}}=1.0$ and $|W_m|$ denotes the number of elements in
that readout matrix.

\paragraph{Optimization.}
Training used AdamW with minibatches of $256$ sequences and at most $100$ epochs.
The visual encoder, GRU, and readout heads used separate optimizer parameter
groups: visual encoder learning rate $2\times10^{-4}$ and weight decay
$10^{-5}$; GRU learning rate $10^{-4}$ and weight decay $10^{-6}$; readout-head
learning rate $10^{-3}$ and no weight decay. Gradients were clipped to a maximum
norm of $1.0$. A ReduceLROnPlateau scheduler monitored the validation objective,
using reduction factor $0.5$, patience $2$, and minimum learning rate $10^{-6}$.
Training stopped early after five epochs without validation improvement. Each
encoder architecture was trained across three run seeds, and the best-validation
checkpoint was saved separately for each encoder architecture and run seed.

\paragraph{Neural-prediction evaluation.}
After training, each model was evaluated separately on each mouse's held-out test
segments. Neural-prediction performance was summarized by Pearson correlation between
predicted and recorded responses, computed both as per-neuron correlations and
as an overall correlation across all predicted and observed response entries. For
the per-neuron metric, correlations were first computed within non-overlapping
100-sample chunks for each neuron, averaged across chunks within each neuron, and
then averaged across neurons. These neural-prediction metrics were used as one axis
of the cross-model analyses.

\subsection{Visual encoder architectures}
\label{app:model_zoo_architectures}

\begingroup
\small
\setlength{\tabcolsep}{5pt}
\renewcommand{\arraystretch}{1.15}

\begin{longtable}{@{}p{0.14\linewidth} p{0.36\linewidth} p{0.43\linewidth}@{}}
\toprule
\textbf{Model label} & \textbf{Visual encoder} & \textbf{Modification} \\
\midrule
\endfirsthead

\toprule
\textbf{Model label} & \textbf{Visual encoder} & \textbf{Initialization / modification} \\
\midrule
\endhead

\midrule
\multicolumn{3}{r}{\emph{Continued on next page}} \\
\endfoot

\bottomrule
\caption{
Visual encoder architectures used in neural-prediction. All encoders
output a $108$-dimensional vector per frame, which is then passed to the same
single-layer GRU with hidden size $512$ and mouse-specific neural readouts.
Standard image-model backbones were modified to accept one-channel grayscale input
and to produce a $108$-dimensional output.
\label{tab:model_zoo_architectures}
}
\endlastfoot

CNN \citep{xu2023multimodal}
&
Original three-block convolutional encoder with channel widths $1\to128\to64\to32$, kernel size $7$, stride $2$, batch normalization, ReLU nonlinearities, dropout, flattening, and a final linear projection.
&
Trained within the neural-prediction pipeline; architecture matches the original Xu et al. encoder except for the shared $108$-dimensional output.
\\

ResNet
&
ResNet-18.
&
First convolution replaced by a one-channel grayscale stem; final fully connected layer replaced by a $108$-output linear layer.
\\

EfficientNet
&
EfficientNet-B0.
&
Convolutional stem replaced for one-channel grayscale input; classifier replaced by a $108$-output linear layer.
\\

MobileNet
&
MobileNet-V2.
&
First convolution replaced for one-channel grayscale input; classifier output replaced by a $108$-output linear layer.
\\

AlexNet
&
AlexNet.
&
First convolution replaced for one-channel grayscale input; classifier modified to map the resulting feature map to $108$ outputs. Final max-pooling operation removed to accommodate the smaller input resolution. 
\\

SqueezeNet
&
SqueezeNet-1.1.
&
First convolution replaced for one-channel grayscale input; final classifier convolution replaced to output $108$ channels before global pooling.
\\

ShuffleNet
&
ShuffleNet-V2 $1.0\times$.
&
Initial convolution replaced for one-channel grayscale input; final fully connected layer replaced by a $108$-output linear layer.
\\

RegNet
&
RegNet-Y-400MF.
&
Stem convolution replaced for one-channel grayscale input; final fully connected layer replaced by a $108$-output linear layer.
\\

\end{longtable}
\endgroup

\subsection{Behavioral readout and feature-extraction details}
\label{app:behavioral_readout_details}

\paragraph{Shared frozen-feature protocol.}
For each task, the trained neural-prediction model was placed in evaluation
mode and all neural-prediction parameters were frozen. We generated synthetic
train, validation, and test stimuli, extracted model features in batches, cached the resulting feature matrices on CPU, and
trained only a task-specific linear probe. Within each run seed, synthetic
train, validation, and test draws were matched across encoder architectures.
Binary tasks used a one-logit linear probe trained with binary cross-entropy,
whereas multiclass tasks used a softmax linear probe trained with
cross-entropy. 
The mouse-specific neural readout heads were not used by task probes.

\paragraph{Task-specific feature convention.}
Static Orientation task used the final visual-encoder output. Each static grating
image was passed directly through the model's visual encoder, yielding a
feature matrix $X\in\mathbb{R}^{N\times108}$. The GRU was constructed when loading the checkpoint but was not used for Orientation feature extraction.

Dynamic Contrast and Motion RDK tasks used GRU features. Dynamic Contrast
extracted the final GRU hidden state from a five-frame clip, giving a feature
matrix $X\in\mathbb{R}^{N\times512}$. Motion RDK first reshaped each
$80$-frame clip into $16$ non-overlapping chunks of five frames, extracted the
last GRU state for each chunk, and then used the final chunk state as the
clip-level feature; this also yielded $X\in\mathbb{R}^{N\times512}$.  Static Orientation and Dynamic Contrast used raw features, whereas Motion RDK applied feature standardization using training-set statistics. We standardized Motion RDK features because the recurrent clip-level features showed substantially larger across-clip variance, and unstandardized linear probes did not reliably converge.

\paragraph{Static Orientation task details.}
Orientation stimuli were full-field sinusoidal gratings rendered at
$60\times80$ pixels. The task used $8$ orientation classes spanning
$[0^\circ,180^\circ)$, with orientations sampled continuously within each bin.
The probe was trained to predict the orientation class from the
$108$-dimensional visual-encoder output. Chance performance was therefore
$1/8$.

Training, validation, and in-distribution test stimuli were sampled from the same nuisance
distribution. Spatial frequency was sampled from
$\{0.02,0.06,0.10,0.14,0.18\}$ cycles per pixel; contrast from
$\{0.70,0.85,1.00\}$; additive noise standard deviation from
$\{0.00,0.04,0.08\}$; Gaussian blur $\sigma$ from
$\{0.00,0.30,0.60\}$; and mean luminance from
$\{0.48,0.50,0.52\}$. Phase was sampled uniformly from $[0,2\pi)$. The compound
out-of-distribution test split used shifted nuisance values: spatial frequency
$\{0.04,0.08,0.12,0.16,0.20\}$ cycles per pixel, contrast
$\{0.55,0.75,0.95\}$, noise standard deviation $\{0.06,0.08,0.10\}$, blur
$\sigma\in\{0.45,0.60,0.80,1.00\}$, and mean luminance
$\{0.46,0.50,0.54\}$.

For each architecture and run seed, probes were trained across a
sample-efficiency sweep with training-set sizes
\[
    N_{\mathrm{train}}
    \in
    \{25,100,250,500,1000,2000,5000,10000,20000\}.
\]
The full training pool contained $20{,}000$ samples, with $2{,}000$ validation
samples, $10{,}000$ IID test samples, and $10{,}000$ OOD test samples. The
primary scalar endpoint for cross-model analyses was accuracy at the largest
training-set size, computed separately for ID and OOD test splits. Normalized
area under the accuracy curve as a function of $\log_{10}(N_{\mathrm{train}})$,
along with chance-adjusted AUC values, was exported as a secondary
sample-efficiency summary.

\paragraph{Dynamic Contrast task details.}
Inspired by \citet{glickfeld2013mouse}, the training clips contained either a flashed Gabor target or a blank noisy
background. Targets were flashed in a fixed visual-field patch within a
$120^\circ\times90^\circ$ visual field mapped to the model's $60\times80$
input grid; training also included nuisance variation in background noise,
carrier spatial frequency, and blur. Target contrasts for probe training were
sampled from
$\{0.05,0.08,0.12,0.18,0.27,0.40,0.60,0.90,1.0\}$. Evaluation used the contrast
grid
$\{0.00,0.02,0.04,0.06,0.08,0.10,0.13,0.17,0.21,0.27,0.34,0.44,0.55,0.70,0.90,1.0\}$.
Positive examples flashed the target on one of the last three frames of the
five-frame clip. Feature standardization was disabled for this task. For each
contrast $c$, balanced accuracy was computed as
\[
    \mathrm{BA}(c) = \frac{1}{2}\left[\mathrm{hit\ rate}(c) +
    \mathrm{correct\ rejection\ rate}\right],
\]
where the correct-rejection rate was estimated from blank trials. The primary
summary was normalized area under the $\mathrm{BA}(c)$ curve.

\paragraph{Motion RDK task details.}
Inspired by \citet{stirman2016touchscreen}, the motion stimuli were $80$-frame random-dot kinematogram clips with dot lifetime
$\tau=20$. The probe was trained on high-coherence conditions
$\{0.95,0.87,0.79,0.71,0.63\}$ and evaluated on the coherence grid
$\{0.00,0.08,0.16,0.24,0.32,0.40,0.48,0.56,0.64,0.72,0.80,0.88,0.95\}$. Feature standardization was enabled: the training-set feature mean and standard
deviation were estimated once per architecture and run seed and then reused for
validation and testing. The primary summary was normalized area under the
accuracy-versus-coherence curve.

\subsection{Emergent tuning and representation-geometry details}
\label{app:emergent_tuning_details}

\paragraph{Feature space.}
Emergent tuning was measured at the final visual-encoder output of each
trained model. For a static image
$\mathbf{x}\in\mathbb{R}^{1\times60\times80}$, the frozen visual encoder
produced a feature vector
\[
    \mathbf{z} = f_\theta(\mathbf{x}) \in \mathbb{R}^{108}.
\]
Each dimension of this vector was treated as a latent feature unit. This final
$108$-dimensional visual-encoder output has no explicit spatial layout, so units
were indexed only by feature dimension. Because these encoder dimensions are
signed model features rather than nonnegative firing rates, all tuning summaries
in this section (orientation selectivity, phase modulation, and contrast-response
fits) use signed-feature variants of the corresponding metrics, as detailed
below.

\paragraph{Stimuli.}
We probed each model with static full-field sinusoidal gratings rendered at the
same spatial resolution as the neural-prediction movies, $60\times80$ pixels. We characterized latent-unit tuning using fixed grids of full-field sinusoidal gratings varying orientation, spatial phase, spatial frequency, and contrast, inspired by classical mouse V1 grating-tuning assays and recent CNN unit-tuning analyses \citep{ringach2016spatial, jang2024improved}.
For pixel coordinates $(x,y)$, orientation $\theta$, spatial frequency $f$ in
cycles per pixel, phase $\phi$, contrast $c$, and mean luminance $\mu=0.5$, the
stimulus intensity was
\[
    I(x,y)
    =
    \mu
    +
    \frac{c}{2}
    \sin\!\left(2\pi f(x\cos\theta + y\sin\theta) + \phi\right),
\]
clipped to the interval $[0,1]$. The main orientation--phase--spatial-frequency
grid used contrast $c=1.0$, $16$ orientations uniformly spaced over
$[0^\circ,180^\circ)$, $16$ phases uniformly spaced over $[0,2\pi)$, and $8$
log-spaced spatial frequencies from $0.01$ to $0.32$ cycles per pixel. Contrast
curves were measured on the grid
\[
    c \in \{0.00, 0.02, 0.04, 0.08, 0.16, 0.32, 0.64, 1.00\}.
\]

\paragraph{Orientation, spatial-frequency, and phase curves.}
For each architecture and run seed, we first evaluated the visual
encoder on the full orientation--phase--spatial-frequency grid. This produced a
response tensor
\[
    R_{o,p,s,u},
\]
where $o$ indexes orientation, $p$ indexes phase, $s$ indexes spatial frequency,
and $u$ indexes the visual-encoder feature dimension. Orientation tuning curves were computed
by averaging over phase and spatial frequency:
\[
    R^{\mathrm{ori}}_{u,o}
    =
    \frac{1}{PS}
    \sum_{p=1}^{P}\sum_{s=1}^{S} R_{o,p,s,u}.
\]
Orientation selectivity was computed after baseline subtraction. Specifically, we defined
\[
    r_{u,o}
    =
    R^{\mathrm{ori}}_{u,o}
    -
    \min_{o'} R^{\mathrm{ori}}_{u,o'},
\]
and computed global orientation selectivity as
\[
    \mathrm{gOSI}_u
    =
    \frac{
    \left|
    \sum_o r_{u,o}\exp(2i\theta_o)
    \right|
    }{
    \sum_o |r_{u,o}|
    }.
\]
The corresponding vector-based preferred orientation was
\[
    \theta^{\ast}_u
    =
    \frac{1}{2}
    \arg\left(
    \sum_o r_{u,o}\exp(2i\theta_o)
    \right)
    \pmod{180^\circ}.
\]

For the subsequent spatial-frequency and phase analyses, we used each unit's
preferred discrete orientation bin from the sampled grid. Spatial-frequency
curves were computed at this preferred orientation by averaging over phase:
\[
    R^{\mathrm{sf}}_{u,s}
    =
    \frac{1}{P}
    \sum_{p=1}^{P} R_{o_u^\ast,p,s,u}.
\]
Spatial-frequency preference was summarized in both sampled-grid and fitted forms. The
sampled-grid estimate was defined as the spatial frequency with maximal response on
\(R^{\mathrm{sf}}_{u,s}\). In addition, we fit each unit's spatial-frequency curve with a
seven-parameter difference-of-Gaussians (DoG) function in log-spatial-frequency space.
For \(x=\log_2(f)\), where \(f\) is spatial frequency in cycles per pixel, the fitted curve was
\[
\hat{R}_u(f) =
R_0
+ A_{\mathrm{exc}}
\exp\left[-\frac{1}{2}\left(\frac{x-\mu_{\mathrm{exc}}}{\sigma_{\mathrm{exc}}}\right)^2\right]
- A_{\mathrm{inh}}
\exp\left[-\frac{1}{2}\left(\frac{x-\mu_{\mathrm{inh}}}{\sigma_{\mathrm{inh}}}\right)^2\right],
\]
with nonnegative excitatory and inhibitory amplitudes. Fits were initialized from five
starting points, and the solution with the largest \(R^2\) on the sampled response curve
was retained. The fitted preferred spatial frequency was defined as the maximum of the
fitted DoG curve on a dense geometrically spaced grid spanning the sampled
spatial-frequency range. A fit was considered usable when it succeeded, produced finite
fit values, and satisfied \(R^2 \geq 0.10\). The primary preferred-SF metadata column used
the fitted preferred spatial frequency; the sampled-grid and fit-only preferred
SF values were also retained as separate metadata columns. The sampled-grid preferred-SF
index was retained for operations requiring an index into the measured response tensor,
including phase and contrast curve extraction.

Phase curves were then computed at each unit's preferred discrete orientation
and preferred discrete spatial frequency:
\[
    R^{\mathrm{phase}}_{u,p}
    =
    R_{o_u^\ast,p,s_u^\ast,u}.
\]
Phase modulation was summarized using the ratio of first-harmonic amplitude to
the absolute DC component of the phase curve,
\[
    \mathrm{F1/F0}_u
    =
    \frac{2|\widehat{R}_{u,1}|}{\max(|\widehat{R}_{u,0}|,\epsilon)},
\]
where $\widehat{R}_{u,k}$ denotes the discrete Fourier coefficient of the phase
curve and $\epsilon=10^{-6}$ prevents division by zero. The preferred phase was
defined as the phase angle of the first Fourier component.

\paragraph{Contrast-response curves.}
Contrast tuning was measured at each unit's preferred orientation and preferred
spatial frequency. To avoid redundant forward passes, units sharing the same
preferred orientation--spatial-frequency pair were grouped together. For each
unique pair, we rendered gratings across the contrast grid and all phase values,
extracted responses from the final $108$-dimensional visual-encoder output, and
averaged over phase:
\[
    R^{\mathrm{contrast}}_{u,c}
    =
    \frac{1}{P}
    \sum_{p=1}^{P}
    R_{c,p,u \mid o_u^\ast, s_u^\ast}.
\]
Each contrast curve was fit with a signed Naka--Rushton function,
\[
    R(c)
    =
    R_0
    +
    R_{\max}
    \frac{c^n}{c^n + C_{50}^{\,n}},
\]
where $R_0$ is the baseline response, $R_{\max}$ is the response gain,
$C_{50}$ is the semisaturation contrast, and $n$ is the exponent. Consistent
with the signed-feature convention noted above, the gain parameter $R_{\max}$
was allowed to be positive or negative. Fits were summarized by $R_0$,
$R_{\max}$, $C_{50}$, $n$, and coefficient of determination $R^2$.

\paragraph{Quality-control flags.}
Quality-control flags were used to identify units with measurable tuning for
summary analyses. A unit was considered responsive if the maximum peak-to-peak
response across its orientation, phase, spatial-frequency, and contrast curves
exceeded $10^{-6}$. Orientation QC required responsiveness, finite gOSI, and an
orientation response range above $10^{-6}$. Phase QC required responsiveness,
finite F1/F0, and a phase response range above $10^{-6}$. Spatial-frequency QC required responsiveness, a spatial-frequency response range above
\(10^{-6}\), a finite preferred spatial frequency, and an acceptable DoG fit with
finite fitted preferred SF and \(R^2 \geq 0.10\). Contrast QC required
responsiveness, a contrast response range above $10^{-6}$, a finite
Naka--Rushton fit, and fit quality $R^2 \geq 0.10$. These thresholds were used
as loose inclusion criteria for descriptive summaries, not as claims that the
corresponding features are biological neurons.

\paragraph{Saved tuning summaries.}
For each architecture and run seed, the emergent tuning pipeline saved a unit-level
metadata table containing one row per visual encoder output feature dimension, including
orientation preference and gOSI, phase F1/F0, sampled-grid and DoG-fitted
spatial-frequency preference and
bandwidth summaries, contrast-fit parameters, and QC flags. The full
orientation, phase, spatial-frequency, and contrast response curves were also
cached. Cross-model analyses used aggregate summaries derived from these
unit-level metadata tables.

\paragraph{Representation eigenspectrum.}
For each trained model, representation geometry was summarized by fitting a
power law to the eigenspectrum of hidden activity. This analysis used a fixed
reference mouse session, \texttt{070921\_J553RT}, sampling 5{,}000 clips from
the concatenated train and test segments. For each clip, we extracted the final
GRU hidden state, standardized each feature dimension
across clips, and applied PCA, retaining up to 256 components capped by feature
dimensionality and sample count minus one. The eigenspectrum exponent $\alpha$
was estimated by linear regression in log--log coordinates over the predefined
rank window 20-200, with $\alpha$ defined as the negative slope of the fitted
line. The same procedure was applied to the visual-encoder output
for diagnostic purposes; only the final-GRU exponent was
used as the representation-geometry column in the cross-model analyses.

\subsection{Cross-model statistical analysis details}
\label{app:cross_model_stats_details}

The cross-model analysis related neural-prediction performance, behavioral
readout performance, emergent tuning, and representation geometry across the
model zoo. Because the neural-prediction objective, recurrent architecture, and
mouse-specific readout structure were held fixed while the visual encoder
architecture varied, cross-architecture variation provided the substrate for
asking how these axes covaried. These analyses were descriptive and
associational rather than causal: they tested whether architectural differences
that affected neural prediction aligned with task readout performance, tuning
structure, or population geometry.

For each run seed $s$ and architecture $a$, the scatter-analysis table
combined four classes of quantities. First, neural-prediction performance was
read from held-out mouse-response prediction summaries, including mean
prediction correlation across the included mouse data. Second, behavioral readout
metrics were read from task-specific summary files: Orientation accuracy at the
largest training-set size, Dynamic Contrast balanced-accuracy AUC, and Motion RDK
accuracy AUC. Third, emergent tuning summaries were computed from the
visual-encoder tuning analysis, including summary orientation selectivity, phase
modulation, and contrast semisaturation. Fourth, representation-geometry
quantities, including the GRU eigenspectrum exponent, were merged for the same
seed and architecture. 

The final statistical table had one row per $(s,a)$ pair before aggregation.
Headline scatter plots used architecture-level means,
\[
    \bar{x}_a = \frac{1}{S}\sum_{s=1}^{S}x_{s,a},
\]
with $S=3$ run seeds. When shown, uncertainty bars reflected
variability across run seeds. Pairwise scatter panels were generated
for a predefined set of cross-axis relationships: neural prediction versus
task performance; neural prediction versus emergent tuning; neural prediction
versus representation geometry; representation geometry versus task performance;
and emergent tuning versus task performance. Each panel reported a fitted linear
trend and an associated correlation or regression statistic.

\subsection{Computational resources}
\label{app:compute_resources}

All models were implemented in PyTorch and run on single-GPU workstations using NVIDIA RTX 4090 GPUs with 24 GB memory and NVIDIA RTX 5060 Ti GPUs with 16 GB memory. Across architectures, the full evaluation pipeline for a single architecture took approximately 1--5 hours to run, including feature extraction, visual readout evaluation, latent-unit tuning analyses, population-geometry analyses, and summary-statistic generation.

\newpage
\section{Model Neural Prediction and Task Performance}

\begin{table}[htbp]
    \centering
    \caption{Model Neural and Task Performance Metrics. Mean $\pm$ cross-neuron SD for neural prediction and Mean $\pm$ cross-seed SD for task performances.}
    \label{tab:model_performance}
    \begin{tabular}{l c c c c}
        \toprule
        \textbf{Model} & \textbf{Neural Prediction} & \textbf{Orientation} & \textbf{Contrast} & \textbf{RDK} \\
        \midrule
        AlexNet    & $0.458 \pm 0.160$ & $0.405 \pm 0.085$ & $0.865 \pm 0.033$ & $0.545 \pm 0.009$ \\
        \citet{xu2023multimodal}        & $0.527 \pm 0.154$ & $0.776 \pm 0.022$ & $0.908 \pm 0.020$ & $0.557 \pm 0.003$ \\
        EfficientNet     & $0.466 \pm 0.166$ & $0.548 \pm 0.043$ & $0.880 \pm 0.026$ & $0.514 \pm 0.014$ \\
        MobileNet  & $0.473 \pm 0.166$ & $0.620 \pm 0.027$ & $0.820 \pm 0.057$ & $0.505 \pm 0.005$ \\
        RegNet     & $0.461 \pm 0.158$ & $0.601 \pm 0.049$ & $0.785 \pm 0.083$ & $0.506 \pm 0.005$ \\
        ResNet     & $0.484 \pm 0.159$ & $0.615 \pm 0.048$ & $0.868 \pm 0.046$ & $0.515 \pm 0.006$ \\
        ShuffleNet & $0.394 \pm 0.161$ & $0.411 \pm 0.052$ & $0.704 \pm 0.032$ & $0.510 \pm 0.006$ \\
        SqueezeNet & $0.533 \pm 0.157$ & $0.614 \pm 0.091$ & $0.906 \pm 0.024$ & $0.562 \pm 0.022$ \\
        \bottomrule
    \end{tabular}
\end{table}

\newpage

\end{document}